\documentclass[aps,amsmath,graphicx,11pt,notitlepage]{revtex4}
\usepackage[dvips]{color,graphicx}
\usepackage{amsmath}
\usepackage{amssymb}

\usepackage[english]{babel}

\def\slasha#1{\setbox0=\hbox{$#1$}#1\hskip-\wd0\hbox to\wd0{\hss\sl/\/\hss}}
\def\slashb#1{\setbox0=\hbox{$#1$}#1\hskip-\wd0\dimen0=5pt\advance
       \dimen0 by-\ht0\advance\dimen0 by\dp0\lower0.5\dimen0\hbox
         to\wd0{\hss\sl/\/\hss}}

\newcommand {\E}[1]{\times 10^{#1}}	

\newcommand{\mc}[1]{\mathcal{#1}}

\begin{document}

\vspace{.5cm}

\begin{center}

{\huge \bf Leptoquarks in FCNC charm decays   }
\thispagestyle{plain}

\vspace{.7cm}

{\large  Svjetlana Fajfer$^{1,2}$ and  Nejc Ko\v{s}nik$^2$  \\}

\vspace{0.5cm}

{\it 1) Department of Physics, University of Ljubljana, 

Jadranska 19, 1000 Ljubljana, Slovenia}

\vspace{0.1cm}

{\it 2) J. Stefan Institute, Jamova 39, P. O. Box 300, 1001 Ljubljana
}

\end{center}

\centerline{\large \bf ABSTRACT}

Recently it was noticed that among many scenarios of new physics
leptoquarks might compensate for the disagreement between the lattice
and experimental results for the charmed strange meson decay
constant. The leptoquarks might modify also the flavor changing
neutral current charm decays.  In this study we investigate
impact of the scalar leptoquark with electric charge $-1/3$ on the
dilepton invariant mass distribution in the $D^+ \to \pi^+\mu^+ \mu^-$
decay and on the branching ratios of the $D^0 \to \mu^+ \mu^-$ using
the existing experimental results.

\vspace{0.5cm}

Recently, a discrepancy between the lattice results and experimental
determination of $f_{D_s}$ in leptonic $D_s \to \ell \nu_\ell$ was
established~\cite{Dobrescu:2008er}. It was
suggested~\cite{Dobrescu:2008er,Benbrik:2008ik} that leptoquarks might
provide for the modification of the short distance part of the
amplitude and account for the difference and at the same time, due to
different helicity suppression, render the $D \to K \ell \nu_\ell$
branching ratio consistent with experiment.
In~\cite{Benbrik:2008ik} authors used the lattice-experiment
discrepancy in $f_{D_s}$, measured branching ratio of semileptonic $D
\to K \ell \nu_\ell$ and some additional assumptions about the mixing
matrices of quarks and leptons to constrain the scalar leptoquark
mediated flavor changing neutral currents, namely the $c \to u \ell^+
\ell^-$.  This resulted in a leptoquark prediction of the $D \to \mu^+
\mu^-$ branching ratio which is close to current experimental
sensitivity.

In this note, we study a possibility to probe the scalar leptoquark
couplings in the $D^+ \to \pi^+ \ell^+ \ell^-$ decay.  We will follow
the approach of~\cite{Fajfer:2007dy,FP-LH}, assuming the
phenomenological Breit-Wigner ansatz for the long-distance resonant
contributions, which by themselves generate branching ratio of around
$2\E{-6}$. Motivated by the new experimental upper
bound~\cite{:2007kg} on the branching ratio
\begin{equation}
\label{cdf}
\mc{B}(D^+ \to \pi^+ \mu^+ \mu^-) < 3.9\E{-6},
\end{equation}
we will find constrains on the scalar leptoquark mediated $c \to u
\mu^+ \mu^-$ Lagrangian. 

Lagrangian of the scalar $\tilde d$ leptoquark in the $(3,1,-1/3)$
representation of the Standard model gauge group, coupled to the
leptons and quarks is~\cite{Dobrescu:2008er,Benbrik:2008ik}
\begin{equation} {\cal L}_{LQ} = \tilde d_\alpha\, \kappa_\ell ( \bar
  \nu_\ell P_R s_\alpha^c - \bar \ell P_R c_\alpha^c) + \tilde
  d_\alpha\, \kappa'_\ell  \bar \ell P_L c_\alpha^c + \mathrm{H.c.},
\label{e1}
\end{equation}
where $\kappa^{(')}$ denotes the effective coupling of second
generation quarks to leptons $\kappa^{(')} = (\kappa^{(')}_e,
  \kappa^{(')}_\mu, \kappa^{(')}_\tau)^T$, $q^c=i \gamma_2 q^*$
are the charge-conjugated quarks, $P_{L,R} = (1\mp \gamma_5)/2$, and
$\alpha$ denotes the color index. This model aims at the resolution of
the puzzle in $D_s \to \ell \nu_\ell$ and in order not to destroy the
consistency between lattice and experiment in $D_d \to \ell \nu_\ell$,
it only couples to quarks of the second generation. For constraints on the first
quark generation, see~\cite{Leurer:1993nr,Leurer:1993em}. The
Lagrangian given in (\ref{e1}) is valid for the quark and lepton
states given in the weak basis. Noticing that the
Cabibbo-Kobayashi-Maskawa matrix is made of the matrices which
diagonalize up and down quarks' mass matrices $V_{CKM}= A_L^{(u)}
A_L^{(d)\dag}$ and Pontecorvo-Maki-Nakagawa-Sakata matrix is made of
$U_{PMNS}= A_L^{(\nu)} A_L^{(\ell)\dag}$, which diagonalize neutrino
and lepton mass matrices, we derived, using Fierz identities, the
following effective Lagrangian for the $c\to u \ell^+ \ell^-$
transition.
\begin{eqnarray} {\cal L}_{\mathrm{eff}}(c \to u \ell^+ \ell^-)=
  \frac{1}{8 M_{\tilde d^2}} \left[C^{L*}_{\ell c} C^L_{\ell u}\,
    (\bar u c)_{V-A}(\bar \ell \ell)_{V-A} +
    C^{R*}_{\ell c} C^R_{\ell u}\, (\bar u c)_{V+A}(\bar \ell \ell)_{V+A} \phantom{\frac{1}{2}}\right.&&\nonumber\\
  +C^{L*}_{\ell c} C^R_{\ell u} \left( \frac{1}{2}\,(\bar u
    \sigma^{\mu \nu}c)
    (\bar \ell\sigma_{\mu \nu} (1-\gamma_5) \ell)  -  (\bar u c)_{S-P}(\bar \ell \ell)_{S-P})\right) &&\nonumber\\
  +\left.C^{R*}_{\ell c} C^L_{\ell u}\, \left(\frac{1}{2} (\bar u
      \sigma^{\mu \nu} c) (\bar \ell\sigma_{\mu \nu}(1+\gamma_5) \ell)
      - (\bar u c)_{S+P}(\bar \ell \ell)_{S+P})\right) \right]&&
  \label{e2}
\end{eqnarray}
In the above expression, effective coupling constants are products of 
leptoquark couplings $\kappa^{(')}$ and fermion mixing matrices:
\begin{subequations}
\begin{align}
C^L &\equiv A_L^{(\ell)} \kappa (A_R^{(u)\dag})_2,\\
C^R &\equiv A_R^{(\ell)} \kappa'(A_L^{(u)\dag})_2,
\end{align}
\end{subequations}
with $(A_{(R,L)}^{(u)\dag})_2$ being the second row of the matrix
$A_{(R,L)}^{(u)\dag}$, and $M_{\tilde d}$ is the mass of the
leptoquark.  In our calculations we do not use additional assumption
on the quark and lepton mixing matrices as it was done
in~\cite{Benbrik:2008ik}. Instead we use the abovementioned general
parameterization.  As noted in~\cite{Benbrik:2008ik}, tensor terms do
not contribute to the bileptonic decay $D\to \ell^+ \ell^-$, while
terms of the type $(V\pm A)\otimes (V\pm A)$ are helicity suppressed
by a factor of $m_\ell$. However, in the $D^+ \to \pi^+ \ell^+ \ell^-$
also terms with tensors contribute without helicity suppression.  We
use the lattice determined hadronic form factors of the $D^+ \to
\pi^+$ transition~\cite{Aubin:2004ej}. For the tensor form factor, we
use the formula, valid in the heavy-quark limit $s(q^2) =
F_1(q^2)/m_D$~\cite{Fajfer:2007dy}.

\begin{figure}[!h]
\begin{tabular}{cp{0.3cm}c}
  \includegraphics[width=0.49\textwidth,angle=-0]{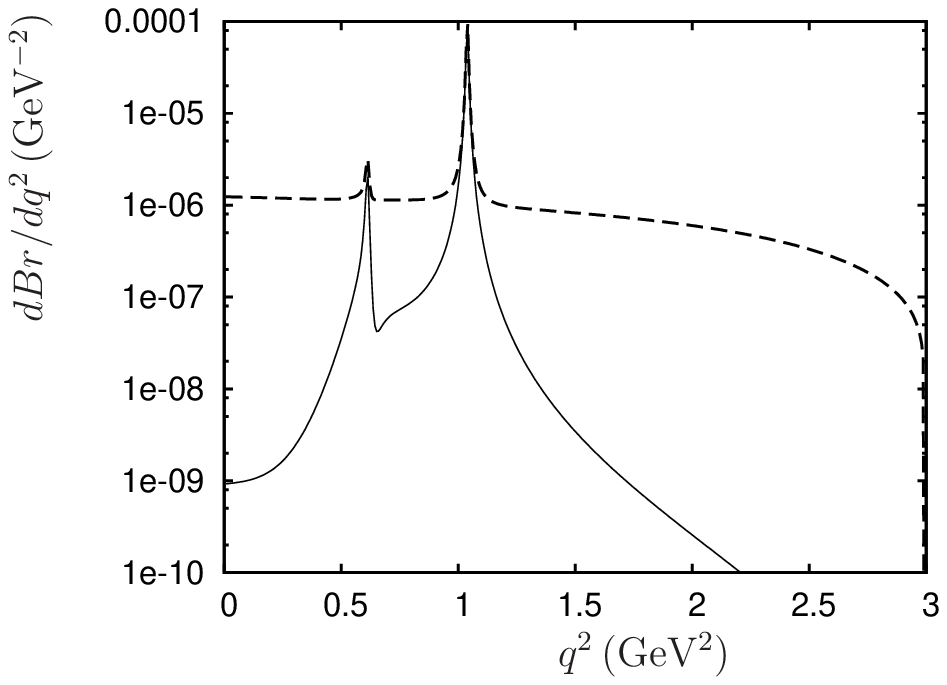} & &
\includegraphics[width=0.49\textwidth,angle=-0]{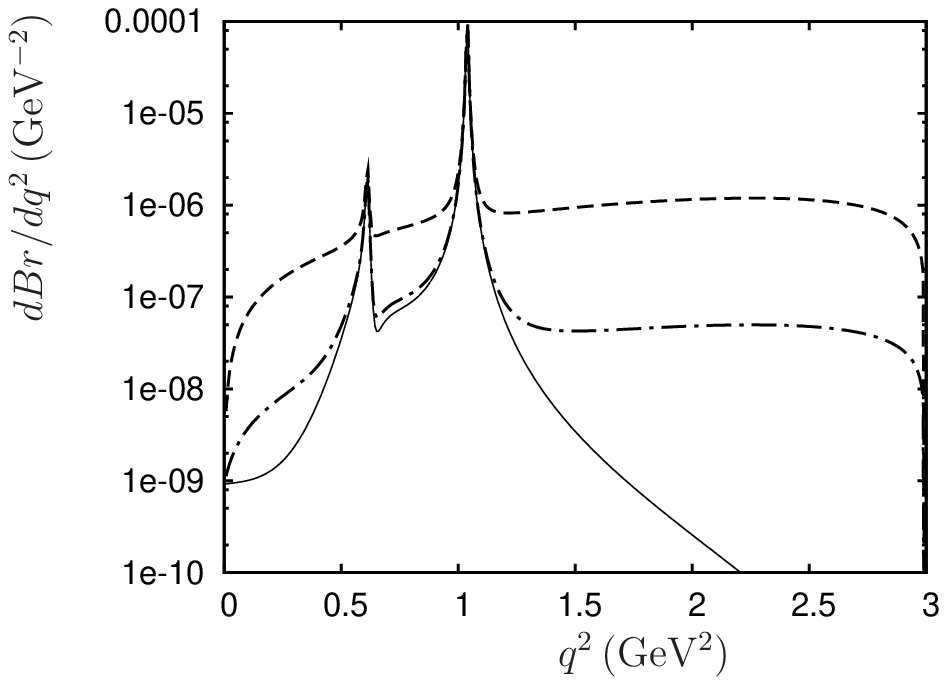}
\end{tabular}
\caption{Spectra of $D^+ \to \pi^+ \mu^+ \mu^-$ decay. Left: Dashed
  line shows saturation of (\ref{cdf}) with the $C^{L(R)}_{\mu c}
  C^{L(R)}_{\mu u}$. Right: Dashed line shows saturation of
  (\ref{cdf}) with the $C^{L(R)}_{\mu c} C^{R(L)}_{\mu u}$, while the
dash-dotted line is the same couplings' contribution, but bounded
from $D^0 \to \mu^+ \mu^-$. Both graphs: Full line is the
long-distance spectrum.}
\label{graphs}
\end{figure}

In the long-distance amplitude we account for the $1^-$ resonances
using the phenomenological Breit-Wigner form (for more details
see~\cite{Fajfer:2007dy, FP-LH}).
\begin{align}
  \label{ampLD}
  \mc{A}^{\mathrm{LD}} = \left[a_\rho \left(\frac{1}{q^2-m_{\rho}^2 +
        i m_{\rho} \Gamma_{\rho}} - \frac{1}{3}
      \frac{1}{q^2-m_{\omega}^2 + i m_{\omega} \Gamma_{\omega}}\right)
    - \frac{a_{\phi}}{q^2-m_{\phi}^2 + i m_{\phi} \Gamma_{\phi}}
  \right]\bar{u}(k_-)\, \slasha{p}\,v(k_+)
\end{align}
Procedure of fitting of the long-distance amplitude to the updated
experimental data~\cite{pdg} is described in~\cite{Fajfer:2007dy} and
we find
\begin{subequations}
\label{fudges}
\begin{align}
  a_\rho &= (2.5 \pm 0.2)\E{-9},\\
  a_\phi &= (4.1 \pm 0.2)\E{-9}.
\end{align}
\end{subequations}
Errors in (\ref{fudges}) are related to experimental errors of $\mc{B}
(D^+ \to \pi^+ \rho)$ and $\mc{B}(D^+ \to \pi^+ \phi)$. The total
differential decay width is a sum of long-distance and leptoquark
contributions, where the former dominates the width in resonant part
of the spectrum and the latter elsewhere. This observation allows us
to neglect the interference term between the long-distance and
leptoquark contributions, as it must be unimportant for the total
decay width~(see Fig.~\ref{graphs}).  The long-distance term generates
partial branching ratio of $(1.8\pm 0.2)\E{-6}$, where the uncertainty
stems from the errors in (\ref{fudges}). In the conservative manner,
we will set the long-distance branching ratio to the minimal allowed
value of $1.6\E{-6}$ and saturate the remaining $2.3\E{-6}$ to the
measured branching ratio (\ref{cdf}) with leptoquark
contribution.
In our study we neglect the mass of the muon, which removes dependence
on the phases of couplings $C^{L,R}$, and arrive at the following
condition.
\begin{equation}
\label{constraint}
  6.07\E{-5} \frac{\left|C^L_{\mu c} C^L_{\mu u}\right|^2 + \left|C^R_{\mu c}
      C^R_{\mu u}\right|^2}{(M_{\tilde d}/\mathrm{TeV})^4}  + 9.15\E{-5} \frac{\left|C^L_{\mu c} C^R_{\mu u}\right|^2 + \left|C^R_{\mu c}
      C^L_{\mu u}\right|^2}{(M_{\tilde d}/\mathrm{TeV})^4} = 2.3\E{-6}
\end{equation}
From (\ref{constraint}) follow the bounds on two distinct combinations
of the leptoquark couplings from the $D^+ \to \pi^+ \mu^+\mu^-$ branching
ratio.
\begin{equation}
\label{bounds3body}
  \frac{|C^{L(R)}_{\mu c} C^{L(R)}_{\mu u}|}{(M_{\tilde d}/\mathrm{TeV})^2} < 0.19, \qquad
  \frac{|C^{L(R)}_{\mu c} C^{R(L)}_{\mu u}|}{(M_{\tilde d}/\mathrm{TeV})^2} < 0.16
\end{equation}
Decay spectra, which saturate (\ref{cdf}), are shown on
Fig.~\ref{graphs} with dashed lines, where
we show the maximal contribution of $C^{L(R)}_{\mu c} C^{L(R)}_{\mu
  u}$ and $C^{L(R)}_{\mu c} C^{R(L)}_{\mu u}$ separately. The former
contribution is proportional to the $F_1(q^2)$ form factor, while the
latter is proportional to the combination of tensor $s(q^2)$ and
$F_0(q^2)$ form factor and also to $q^2$. The latter couplings are
present also in the bileptonic decay $D^0 \to \mu^+ \mu^-$, for which
the branching ratio (with $m_\mu$ again set to zero and $f_D$ taken
from \cite{:2008sq}) is
\begin{equation}
  \label{width2body}
  \mc{B} (D^0 \to \mu^+ \mu^-) = \tau_{D_0} \frac{f_D^2 m_{D_0}^5}{256 \pi m_c^2}
 \frac{|C^L_{\mu c} C^R_{\mu u}|^2+|C^R_{\mu c} C^L_{\mu u}|^2}{M_{\tilde d}^4}.
\end{equation}
From experimental result $\mc{B} (D^0 \to \mu^+ \mu^-) <
5.3\E{-7}$~\cite{Bussey:2008eg} we obtain the bound
\begin{equation}
  \label{bounds2body}
  \frac{|C^{L(R)}_{\mu c} C^{R(L)}_{\mu u}|}{(M_{\tilde d}/\mathrm{TeV})^2} < 0.032,
\end{equation}
which is 1 order of magnitude more severe than the bound from $D^+ \to
\pi^+ \mu^+ \mu^-$~(\ref{bounds3body}). From (\ref{bounds2body}) we
predict the maximal contribution of $C^{L(R)}_{\mu c} C^{R(L)}_{\mu
  u}$ to $\mc{B}(D^+ \to \pi^+ \mu^+ \mu^-)$ to be $9.4\E{-8}$
(Fig.~\ref{graphs}, dash-dotted line) which is almost impossible to
detect due to the resonant branching ratio pollution of $(1.8\pm
0.2)\E{-6}$. However, the $C^{L(R)}_{\mu c} C^{L(R)}_{\mu u}$ can only
be probed by measuring the spectrum of $D^+ \to \pi^+ \mu^+ \mu^-$,
where an enhancement of about 3 orders of magnitude with respect to
the resonant amplitude is expected at low $q^2$ (dashed line in
Fig.\ref{graphs}, left).

We estimated the scalar $Q=-1/3$ leptoquark contribution to $D^+ \to
\pi^+ \mu^+ \mu^-$ and $D^0 \to \mu^+ \mu^-$ branching ratios.  The
3-body decay mode is sensitive to both $C^{L(R)}_{\mu c} C^{R(L)}_{\mu
  u}$ and $C^{L(R)}_{\mu c} C^{L(R)}_{\mu u}$, and we have shown that
with additional input from the bileptonic decay, one can use the
measured branching ratio of $D^+ \to \pi^+ \mu^+ \mu^-$ to probe
directly the leptoquark couplings that are otherwise inaccessible to
$D^0 \to \mu^+ \mu^-$. Further experimental study of $D^+ \to \pi^+
\mu^+ \mu^-$, especially in the low $q^2$ region, together with
improving measurements of $D^0 \to \ell^+ \ell^-$, $D \to K \nu \ell$
and $D^0-\bar{D}^0$ mixing~\cite{Golowich:2007ka} can help decide
whether leptoquarks can solve the $f_{D_s}$ puzzle.






\begin{acknowledgments}
  We thank Bo\v stjan Golob for useful discussions on the experimental
  side of this work.  This work is supported in part by the European
  Commission RTN network, Contract No. MRTN-CT-2006-035482
  (FLAVIAnet), and by the Slovenian Research Agency.
\end{acknowledgments}

\bibliography{refs}

\end{document}